\documentclass[aps,prl,reprint,amsmath,amssymb,floatfix]{revtex4-2}

\usepackage[T1]{fontenc}
\usepackage{mathrsfs}
\usepackage{gensymb}
\usepackage{amsthm}
\usepackage{graphicx}
\usepackage{xcolor}
\usepackage{bm}
\usepackage[caption=false]{subfig}
\usepackage{verbatim}
\usepackage{siunitx}
\usepackage{tabularx}
\usepackage{booktabs}
\usepackage{multirow}
\usepackage{mathtools}

\usepackage{bibunits}
\defaultbibliographystyle{apsrev4-2}

\usepackage{hyperref}

\begin{document}

%\begin{bibunit}[apsrev4-2]
\title{Unraveling hidden Raman modes in $\alpha$-MnTe: Effects of long-range dipole-dipole interactions on phonon bands}
\author{Susmita Jana$^{1,2}$}
\email{jsusmita.research@gmail.com}
\author{Rajib Sarkar$^{3}$}
\author{Dhavala Suri$^{3}$}
\author{ B.R.K Nanda$^{1,2}$}
\email{nandab@iitm.ac.in}
\affiliation{$^1$Condensed Matter Theory and Computational Lab, Department of Physics, IIT Madras, Chennai, 600036, Tamil Nadu, India}
\affiliation{$^2$Center for Atomistic Modelling and Materials Design, IIT Madras, Chennai, 600036, Tamil Nadu, India}
\affiliation{$^3$Centre for Nanoscience and Engineering, Indian Institute of Science, Bengaluru, Karnataka 560012, India} 

%\date{September 2026}
\begin{abstract}
In this study, we address the seemingly unresolved question of disparity between the experimentally and theoretically observed Raman spectra for the altermagnetic $\alpha$-MnTe. We establish that the consideration of long range (LR) dipole-dipole interaction is crucial to accurately estimate the phonon band structure and thereby capturing each of the experimentally observed Raman signatures. The LR correction brings about substantial changes to the high-frequency optical modes and their group-theoretical representations. Specifically, for the highly debated Raman peak around 175 cm$^{-1}$, alongside the Raman silent B$_{1u}$ mode, the LR corrections introduce Raman-active longitudinal optical A$^{\prime}$ and A$_1$, and transverse optical A$_1$ modes. Overall, the calculated Raman spectra suggest that the intensity of these modes is weak in the pristine system, while certain prescribed distortions can increase their intensity. The Raman peak around 140 cm$^{-1}$, as observed experimentally, is identified to be emerging from a higher-order Raman process due to zone off-center contributions, while the other debated 120 cm$^{-1}$ peak has both the zone center and off-center contributions.

\end{abstract}
\maketitle

%\section{Introduction}
Over the last few years the insulating $\alpha$-MnTe (henceforth referred as MnTe) has been intensely investigated as it exhibits non-relativistic spin splitting of the bands, a characteristic feature of altermagnetism\cite{vsmejkal2022beyond}, and finite anomalous Hall conductivity\cite{sarkar2026anomalous, li2022antiferromagnetic}. Alongside the electronic and magnetic structure\cite{lee2024broken, chilcote2024stoichiometry}, the experimental Raman characterization, one of the fundamental sample characterization tools, has opened up a new fascinating research problem involving this compound. Broadly, the experimental results are segregated into two categories, one with the presence of a Raman peak at 175 cm$^{-1}$\cite{kluczyk2024coexistence, zhang2020spin, wu2026domains, uykur2026revisiting} and the others which do not observe such peaks\cite{szuszkiewicz1997temperature, azam2026origin, bhardwaj2026nature175180cm1raman}.

The MnTe films that are grown on InP (111) substrate, as reported by Zhang \textit{et. al.}\cite{zhang2020spin} show Raman peaks at 122.5 cm$^{-1}$, 140.6 cm$^{-1}$, 175.7 cm$^{-1}$, 268.8 cm$^{-1}$, and 345.8 cm$^{-1}$. 
%The peak at 175.7 cm$^{-1}$ is assigned to the E$_{2g}$ mode.
There are past studies that did not report the Raman peak at around 175 cm$^{-1}$. For example, two of the earlier Raman measurements\cite{szuszkiewicz1997temperature, szuszkiewicz2014raman} on $\mu$m thick MnTe films, carried out at low temperature ($<$ 30 K) do not exhibit excitation at 175 cm$^{-1}$. However, these studies report Raman excitation in the range 195 - 215 cm$^{-1}$. Another recent experimental work reveals that the MnTe films grown on GaAs (111) substrate show Raman peaks at 90 cm$^{-1}$, 120 cm$^{-1}$, and 140 cm$^{-1}$\cite{azam2026origin}. Interestingly, the air exposed samples in this study did not exhibit the 175 cm$^{-1}$ peak, even though one would expect oxidation and Te excess surface in them. In a recent experimental study on MBE grown MnTe film on Si(111) substrate, we noticed mixed observations. The pristine spot showed Raman intensity peaks around 90-105 cm$^{-1}$, 120 cm$^{-1}$, and 140 cm$^{-1}$. The spot burned with laser irradiation showed these peaks as well as the peak around 175 cm$^{-1}$.\cite{bhardwaj2026nature175180cm1raman}. We attribute the latter to the signature of distorted MnTe or of extrinsic MnTe$_2$ which has a well established A$_g$ peak at 175 cm$^{-1}$\cite{palasyuk2026paramagnetic}. 

 In the case of CVD grown bulk MnTe single crystals, polarization dependent Raman measurements show peaks at 120 cm$^{-1}$, 140 cm$^{-1}$, 175 cm$^{-1}$, and 270 cm$^{-1}$\cite{wu2026domains}. In the pristine system, the $\Gamma$-centered optical phonon mode, $B_{1u}$, around 175 cm$^{-1}$ is Raman silent. The authors propose that a probable distortion, reducing the symmetry of the $D_{6h}$ to $D_{3h}$ point group, converts the $B_{1u}$ mode to $A1^{\prime}$ and makes it Raman active. Similarly, another study\cite{uykur2026revisiting} on bulk single crystal MnTe also reports the Raman peak at 175 cm$^{-1}$. However, in the absence of any symmetry lowering, the authors here assigned the peak to the possible presence of residual MnTe$_2$. Differing with the above, another report on bulk MnTe showed 175 cm$^{-1}$ Raman peak and assigned it to E$_{2g}$ mode. \cite{kluczyk2024coexistence} 
 %(see the supplementary material of this paper).

 Concurrent to the experimental studies, several first-principles calculations are carried out to obtain the phonon modes and thereby the Raman signatures in MnTe. Thapa \textit{et. al.} ruled out the possibility that the E$_{2g}$ mode is responsible for Raman excitation at 175 cm$^{-1}$. Instead, they proposed it as an outcome of a plasmonic (electronic excitations) origin, generated due to self hole doping. Specific to 175 cm$^{-1}$ peak, Wu \textit{et. al.}\cite{wu2026domains} proposed that, a plausible symmetry lowering distortion in highly symmetric MnTe can result in a leakage of 175 cm$^{-1}$ mode. The theoretical calculations of Raman active modes reported in \cite{azam2026origin} did not identify the concerned feature while accounting the in-plane tensile strain induced by the GaAs substrate and its temperature dependence. A DFT+U study by Uykur \textit{et al.} \cite{uykur2026revisiting} in MnTe, with varying U values and different exchange-correlation functionals, reveals the presence of two closely spaced Raman active modes in the range $\sim$86-102 cm$^{-1}$ and no other Raman active modes. Interestingly, the experimental study in this work reports the existence of two other strong modes at 120 cm$^{-1}$ and 140 cm$^{-1}$. The latter are believed to be non $\Gamma$-centered modes which are activated by higher order Raman processes. This feature has not been substantiated by any other work.

The contradictions among the varied reports, as mentioned above, have only increased the curiosity to address this seemingly open question. In this work, we demonstrate the presence of all the experimentally observed Raman peaks and proposed reasons on why some of the experiments may be missing them. We also show that the long-range dipole-dipole interaction, not employed in earlier studies, significantly modifies the phonon band structure and creates many low-intense phonon modes including the ones around 175 cm$^{-1}$. We further show that the higher-order Raman process with zone off-center contribution generates the rarely observed 140 cm$^{-1}$ peak. 

The harmonic forces required to analyze the vibrational modes of MnTe were calculated within density functional theory (DFT) as implemented in VASP. The interatomic forces were calculated using the PAW method with the PBE-GGA exchange-correlation functional. The crystal structure was fully relaxed
%with respect to both lattice parameters and atomic positions 
while retaining the A type N\'eel order of this altermagnet. To account for the effect of correlations, the DFT+U method is employed through the Dudarev approach\cite{dudarev1998electron}. A symmetric 2$\times$2$\times$2 supercell is used to incorporate the vibrations of the atoms.
%at the primitive cell boundary. 
After calculating a set of converged forces using density-functional perturbation theory (DFPT), the dynamical matrices were obtained on a 9$\times$9$\times$3  $k$-mesh and, the same set of convergence parameters were used as in the ground-state calculations. The dynamical matrix is solved using Phonopy\cite{phonopy-phono3py-JPCM} to obtain the phonon eigenvalues and eigenvectors.
%were subsequently calculated as a solution of the dynamical matrix using Phonopy\cite{phonopy-phono3py-JPCM}.

\begin{figure}[]
    \centering
    \includegraphics[width=\columnwidth]{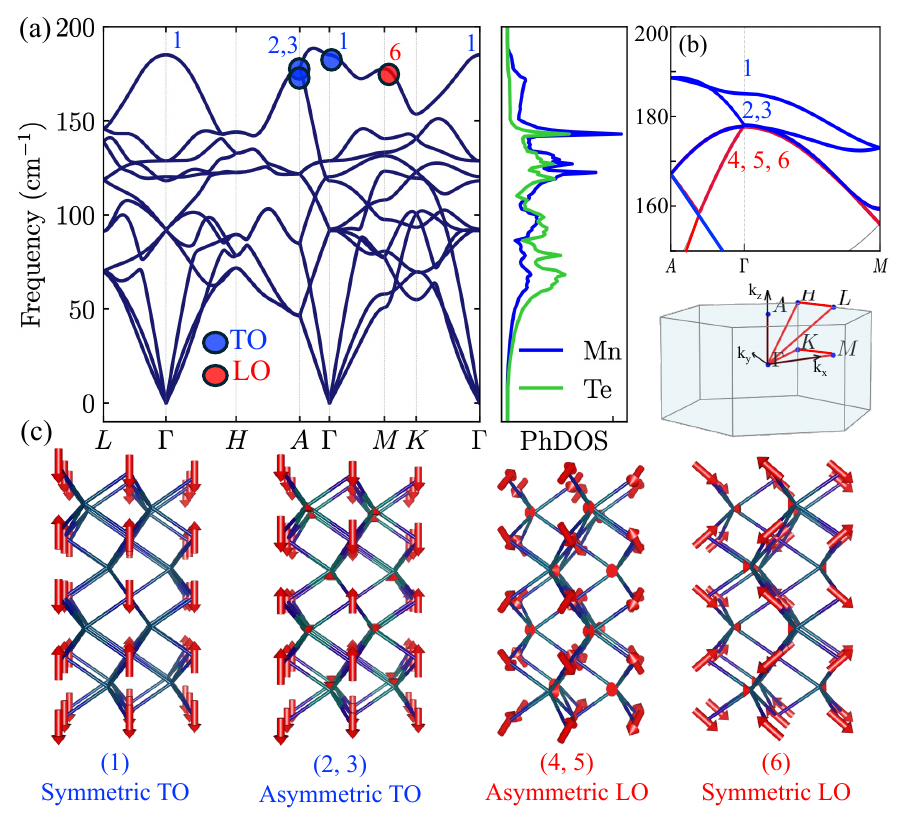}
    \caption{(a) Phonon band structure and atom projected phonon density of states (PhDOS) of bulk MnTe without LR-correction. The high-symmetry points connecting the q-path is indicated in the Brillouin zone (right). The blue and red points on the spectra at $\Gamma$, A and M points indicate three independent transverse optical (TO) and, one longitudinal optical (LO) vibrations observed in MnTe system. (b) The phonon band structure for the 2$\times$2$\times$2 supercell of MnTe from which the bands are unfolded to represent the primitive unit cell band structure, as in Fig. (a). The blue and red bands are only shown for the high frequency TO and LO optical phonons, respectively. Once unfolded, the modes, 2-6 do not lie at $\Gamma$ and are mapped to other high symmetry points of the primitive unit of MnTe. We mapped (2, 3) and (6) belong to A and M, respectively. The vibrations corresponding to these identified LO and TO modes are shown in panel (c). We have classified the vibrations to be symmetric if the neighboring Mn atoms exhibit identical displacement amplitudes and directions, and asymmetric when their displacement patterns differ.}
    \label{prim+super}
\end{figure}

The DFT+U (U = 4 eV) calculated phonon band structure and atom-projected phonon density of states of MnTe are shown in Fig. \ref{prim+super}(a). 
%This band structure is in agreement with the ones reported in the literature\cite{azam2026origin}. 
This band structure is obtained by unfolding the phonon bands of the supercell 
%calculations performed using a $2\times2\times2$ supercell 
onto the Brillouin zone (BZ) of the primitive unit cell and it is in agreement with the literature\cite{azam2026origin}. For comparison, the folded supercell phonon bands in the region of interest (150–200 cm$^{-1}$) are shown in Fig. \ref{prim+super}(b), where we observe six high-frequency solutions of the dynamical matrix lying at $\Gamma$. The highest optical mode (1), at 185.09 cm$^{-1}$, corresponds to a symmetric transverse optical (S-TO) vibration. The two doubly degenerate modes (2, 3), at 178.05 cm$^{-1}$, correspond to asymmetric transverse optical (A-TO) vibrations. The three degenerate longitudinal optical (LO) modes occur at 177.74 cm$^{-1}$ and consist of two asymmetric LO (A-LO) and one symmetric LO (S-LO) modes. The displacement patterns of these modes are represented by vectors attached to the atomic sites in Fig. 1(c). When these supercell phonon modes are unfolded onto the BZ of the primitive unit cell, the S-TO mode remained at $\Gamma$, whereas the two A-TO modes are mapped onto the high-symmetry point A, as marked in Fig. \ref{prim+super}(a). Among the three LO modes, only the S-LO mode appears at the M point, while the two A-LO modes could not be discerned.
%in the unfolded phonon bandstructure. 
Since all of them are located in the region of interest (175 cm$^{-1}$), it is important to analyze these vibrations along with the zone-center one to provide a broader understanding of the high frequency vibrational response of MnTe.

There is a notable difference in the phonon band structure, particularly the optical phonon bands, of Fig. \ref{prim+super}(a) from the one shown by Wu \textit{et al.} \cite{wu2026domains}, which suggests that supercell convergence is a necessary sanity check. In Fig.S1 of the Supplementary Material(SM)-I, we have shown the evolution of the phonon band structure with the supercell size and this implies that the band structure corresponding to the 3$\times$3$\times$3 supercell fits well with this literature. The supercell convergence test also tells us that the flat band observed for the primitive unit becomes oscillatory as we increase the supercell. %\textcolor{red}{This is due to the fact that  with increase in the supercell size, smaller phonon wavevectors (First BZ becomes even smaller) are sampled to describe the same vibrations.}
The other interesting observation is the presence of an unusually steep phonon dispersion along $\Gamma$-A for supercells 2$\times$ 2 $\times$ 2 and 3$\times$ 3 $\times$ 3. Such an abrupt dispersion near $\Gamma$ can be linked with the finite size truncation of the vibration space. In fact, it is known that the estimation of accurate phonon modes requires very large supercells. Particularly, in the case of semiconductors and insulators, incomplete electronic screening in the case of smaller supercells can lead to inaccurate estimation of the interatomic force constants which become a prominent factor in the case of polar systems like NaCl, MgO, KCl etc.\cite{togo2022mode, coiana2024breakdown}. In such systems, the finite size truncation can lead to Gibb's oscillation \cite{gonze1997dynamical, phonopy-phono3py-JPCM}.  As it is practically impossible to compute an infinitely large supercell for the calculations, alternatively, one can consider the non-analytic correction (NAC) using the Born effective charges and dielectric tensor proposed by Gonze et. al.\cite{gonze1997dynamical} to account for the long-range (LR) behavior. The Born effective charges, listed in Tables S1 and S2 of SM-II, are found to be substantial for MnTe, and therefore the LR dipole interaction correction is required to calculate the phonon band structure. The NAC incorporated phonon band structure of MnTe is shown in the lower panel of Fig. S1 of SM, which not only implements significant corrections to the optical phonon modes, but also shows convergence with a 2$\times$ 2 $\times$ 2 supercell. A detailed analysis of LR interaction induced changes in MnTe using mode dependent Born effective charges is provided in SM-II.
\begin{figure}[t]
    \centering
    \makebox[\columnwidth][c]{%
        \includegraphics[width=1.15\columnwidth]{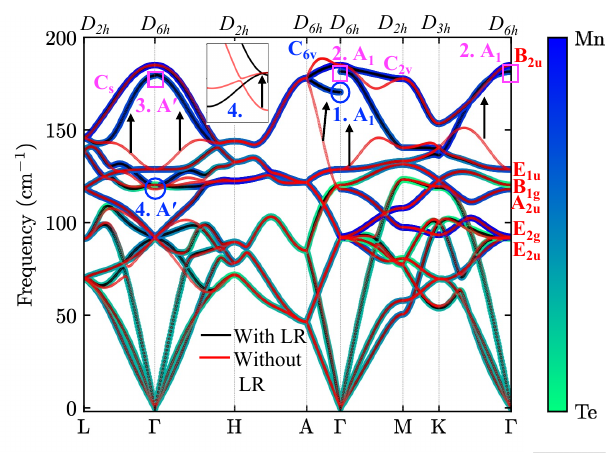}
        }
    \caption{The comparison of phonon band structure without LR interaction (red solid lines) and with LR interaction (color projected Mn and Te contributions). The inclusion of LR lifts the degeneracy in high frequency optical modes along certain symmetry paths and the changes are indicated through arrows.}
    \label{SR+LR}
\end{figure}

\begin{figure*}[t]
    %\centering
    \includegraphics[width=\textwidth]{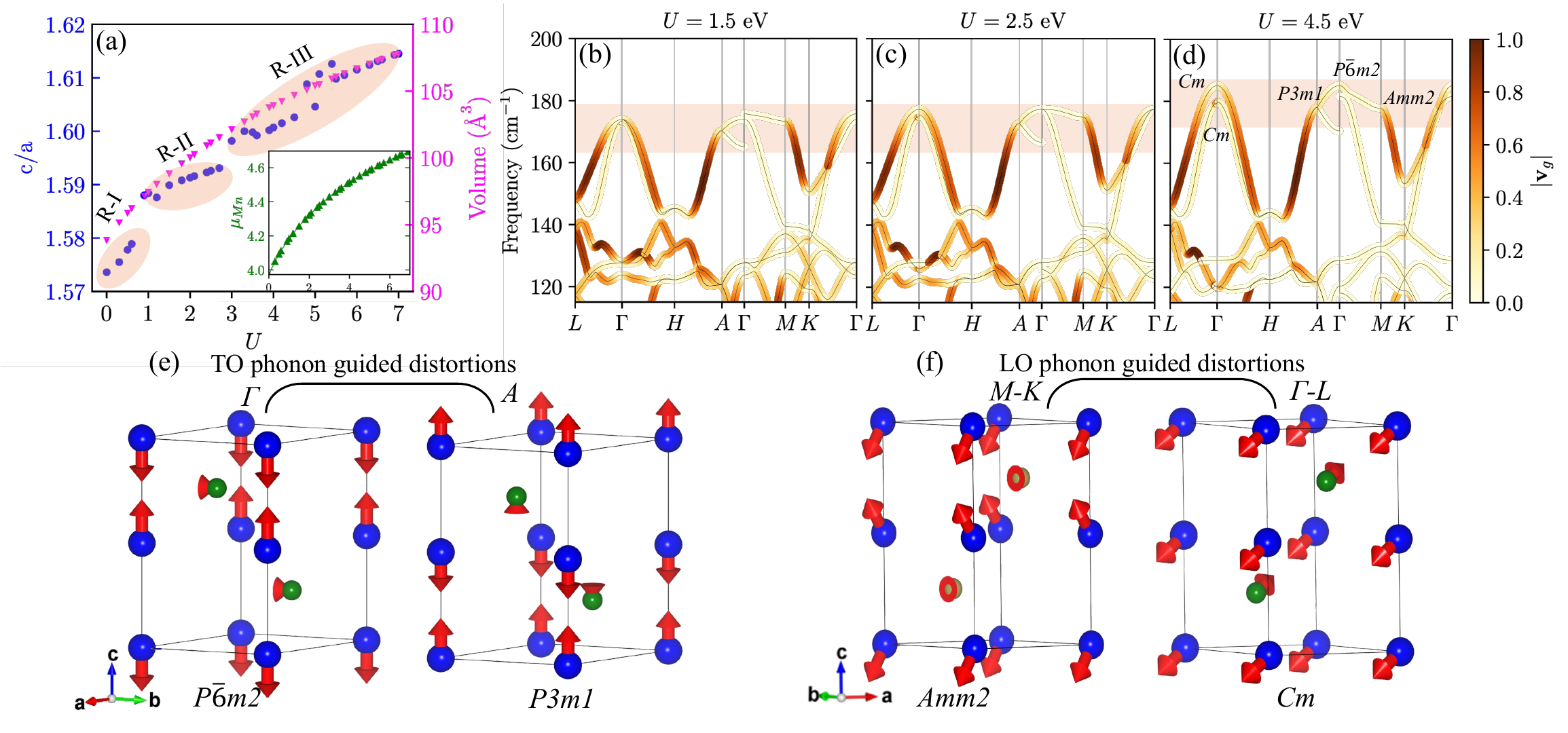}
    \caption{(a) The $c/a$ ratio and the primitive-cell volume as a function of the effective Hubbard parameter (U). Normalized group-velocity ($v_g$) maps overlaid with the LR-corrected phonon band structures for (b) U = 1.5 eV, (c) 2 eV, and (d) 4 eV. The results indicate the U-dependent behaviour of the high frequency optical phonon branches. Schematic illustrations of the structural distortions generated by displacements along the corresponding (e) TO, and (f) LO phonon eigenvectors of MnTe. The space groups of the resulting distorted structures are indicated.} %\textcolor{red}{(I) Place the magnetization curve in the inset. (II) Figures are not labeled. (III) The axes a, b, c are invisible. I think you need not show them. (IV) Question: How will one distinguish which one is LO and which one is TO? (V) In the figure itself, indicate regime-1, 2, and 3.} }
    \label{gv_dist}
\end{figure*}

To further elucidate the distinct features introduced by the LR dipole-dipole interactions correction, in Fig. \ref{SR+LR}, we represent the atom projected phonon band structure and compared them with the bands obtained without the LR corrections (red solid lines). %The relevant group theoretical representations to the optical modes are also assigned according to their point group symmetry.
We would also like to add that, since the corrected optical modes are not well-defined at $\Gamma$ due to directional dependence, we assigned their symmetry by comparing their characters with the appropriate irreducible representations along the corresponding high symmetry q-path in the BZ. The important changes by the LR correction, indicated by arrows in Fig. \ref{SR+LR}, are as follows. 
(i) The steep dispersion of the TO phonons that rises from 118 cm$^{-1}$ along $\Gamma$ - A disappears and instead a new branch of TO emerges from A at a degenerate point (178 cm$^{-1}$) and reaches 170.4 cm$^{-1}$ at $\Gamma$. This TO mode, marked by a blue circle, follows the $C_{6v}$ symmetry of the $\Gamma$-A path and is assigned the $A_1$ mode.
(ii) The LO branch that starts at 140 cm$^{-1}$ at M and reaches 128.8 cm$^{-1}$ $\Gamma$ rises to 181.6 cm${-1}$. A similar one is observed along K-$\Gamma$. Both are marked by a pink rectangle and are also assigned the $A_1$ mode based on the $C_{2v}$ symmetry along the $\Gamma$-M (K) path.
%A second splitting occurs between the degenerate LO modes near 128.8 cm$^{-1}$ along M-$\Gamma$ and $\Gamma$-K direction. 
(iii) Focusing on the path along L-$\Gamma$-H, we find that a branch of LO in the range 128 cm$^{-1}$ changes the curvature from downward to upward, as indicated by the arrows, and reaches 179 $\pm \delta$ cm$^{-1}$ with a discontinuity at $\Gamma$. This corrected LO mode (pink rectangle) follows the $C_s$ symmetry of the $\Gamma$-L (H) path and is assigned the $A^{\prime}$ mode.
%Another LO splitting is observed along $\Gamma$-L(H) direction as well.
(iv) One of the TO modes at $\Gamma$, marked with a blue circle and amplified in the inset, shifts from 118 to 120 cm$^{-1}$. The LR correction lifts phonon degeneracy and modifies the harmonic vibrational energy landscapes around those above mentioned frequency values. Modes III and IV exhibit a significant increase in their effective force constants (see Fig. S2, SM-III). This mode is assigned with $A^{\prime}$ and has a significant role in modifying the Raman spectra, as discussed later. Vibrational modes are assigned to appropriate irreducible representations by performing the point group dictated symmetry operations to their phonon eigenvectors and comparing them with the corresponding character tables (see SM-IV for details).

\begin{table}[t]
\caption{\label{lr_list} Raman modes with their group theoretical representations. The new Raman modes after the inclusion of LR corrections are highlighted in color. As observed, without LR interaction, phonon modes were either silent or, infrared active. After the LR driven splitting, new Raman active phonon modes are found and therefore additional peaks emerge in the frequency range of 120 -190 cm$^{-1}$.}
\centering
\footnotesize
\setlength{\tabcolsep}{3.5pt}
\renewcommand{\arraystretch}{1.12}
\begin{ruledtabular}
\begin{tabular}{cccc}
Frequency & Vibrational & Mode & Activity \\
(cm$^{-1}$) & signature & representation & \\
\hline
185.03     & Mn: TO ($\Gamma$)     & $B_{2u}$ ($D_{6h}$) & S  \\
\textbf{181.56}    & \textbf{Mn: LO ($\Gamma$-M)}  & $\boldsymbol{A_1 (C_{2v})}$   & \textbf{R}  \\
\textbf{179.36}     & \textbf{Mn: LO ($\Gamma$-L)}  & $\boldsymbol{A^\prime (C_s)}$        & \textbf{R}  \\
\textbf{170.42}     & \textbf{Mn: TO ($\Gamma$-A)}  & $\boldsymbol{A_1 (C_{6v})}$   & \textbf{R}  \\
128.74 (2) & Mn: LO ($\Gamma$)     & $E_{1u}$ ($D_{6h}$) & IR \\
120.35     & Te: TO ($\Gamma$)     & $B_{1g}$ ($D_{6h}$) & IR \\
\textbf{120.22}     & \textbf{Mn: TO ($\Gamma$-L)}  & $\boldsymbol{A^{\prime} (C_s)}$       & \textbf{R}  \\
118.13     & Mn: TO ($\Gamma$)     & $A_{2u}$ ($D_{6h}$) & S  \\
92.35 (2)  & Mn: LO ($\Gamma$)     & $E_{2g}$ ($D_{6h}$) & R  \\
91.58 (2)  & Te: LO ($\Gamma$)     & $E_{2u}$ ($D_{6h}$) & S  \\
Acoustic (3) & $\Gamma$           & --                 & -- \\
\end{tabular}
\end{ruledtabular}
\end{table}

Table \ref{lr_list} lists the existing and the new (bold) Raman modes after the LR correction. Their group theoretical representations and optical activities are also mentioned. Interestingly, we find that all the four new modes, A$_1$ (TO), A$_1$ (LO), A$^{\prime}$ (LO), A$^{\prime}$ (TO), denoted in Fig. \ref{SR+LR}, are Raman active according to the selection rules of their respective path symmetries and they emphasize the necessity of the LR dipole-dipole interactions to explain the origin and robustness of the Raman response measured in MnTe.

The strong correlation has a profound effect on the electronic structure of MnTe\cite{chernov2025electronic, kang2026strong}. With a change in the electronic structure,  the mean-field,  responsible for the lattice vibration, also changes. Therefore, it is prudent to examine the effect of the correlation on the vibrational response of MnTe. To achieve this, we performed structural relaxation as a function of U, and the results are summarized in Fig. \ref{gv_dist}. From Fig. \ref{gv_dist}(a), we see that both lattice volume and the $c/a$ ratio increase progressively with increasing U, indicating an evolution of the equilibrium structure. We also note that with an increase in c/a, the system achieves a saturated local magnetization consistent with previous reports\cite{carlisle2025tuning, chernov2025electronic}.

Subsequently, the LR interaction included the phonon band structure for each value of U, using the corresponding relaxed lattice parameters are calculated, and the results are shown in Fig. \ref{gv_dist}(b). %for three U values (1.5 eV, 2 eV, and 4 eV).
From the figure we gather that there are three major changes around the zone-center with increase in U. First, the optical modes gradually shift toward higher frequencies with increasing U. Second, concerning the bands along A-$\Gamma$-M, we find that the relative ordering of the top LO and top TO bands changes with increase in U, resulting in a LO–TO crossover (marked by a shaded region). Third, a distinct splitting occurs between Mn and Te dominated vibrational branches around 90 and 120 cm$^{-1}$ (not shown here). These changes imply that the vibrational response of MnTe is strongly coupled to the U-dependent structural changes. 
Based on the distinct structural and vibrational trends, U dependent behavior can be broadly divided into three characteristic regimes, namely, R-I: 0 $\le$ U$\le$ 1.8 eV; R-II: 2 $\le$ U $\le$ 2.8 eV; and R-III: 3.0 $\le$ U $\le$ 7.0 eV, as indicated in Fig. 3(a).

As seen from Fig. \ref{gv_dist}(b), for the top optical branches, the dispersion gradually steepens with increasing U. To characterize it further and to shed light on the directional behavior of these high frequency vibrations, we examine the phonon group velocity, $v_g$, along the three independent high-symmetry directions, $\Gamma$-A, $\Gamma$-M, and $\Gamma$-L, respectively (see SM-IV for more details). 
% Since $v_g$=0, exactly at the zone-center for the optical modes, we focus on the dispersion gradient in the vicinity of $\Gamma$ for the two optical branches. 
For the first and second correlation regimes, the highest LO, TO modes both exhibit negligible $v_g$ along A-$\Gamma$-M. However, in the third regime, the $v_g$ for TO along $\Gamma$-M (25.51 km/s) becoming approximately five times larger than that along $\Gamma$-A (5.97 km/s). In contrast, along $\Gamma$-L, $v_g$ is consistently the largest both for TO and LO modes across all U ranges. In fact, we find that the $v_g$ of the TO (255.73 km/s) and LO (2613.92 km/s) modes along $\Gamma$-L are nearly two and three orders of magnitude larger, respectively, than those along $\Gamma$-M. The directional dependence of $v_g$, reveals that the progression of the zone-center modes associated with specific distortions can be a prominent pathway to understand the vibrational fingerprint of MnTe.%(\textcolor{red}{What does it mean and why is it so important?})

Next, we examine the structural alterations arising from the high frequency vibrations and subsequent Raman response. This is achieved by ``freezing" the phonon eigenvectors of the six modes (three LO and three TO) and analyzing the resulting symmetry lowered structures. Each mode produces a unique distortion in MnTe and lowers the parent space group (SG) symmetry $P6_{3}/mmc$ (194) to a corresponding maximal subgroup symmetry, as indicated in Fig. \ref{gv_dist}(c). The distortion directions can be visualized from the lower panel of Fig. \ref{gv_dist}.
 The freezing of the S-TO mode produces a structure with SG $P\bar{6}m2$ (187), whereas the A-TO mode produces a structure with SG $P3m1$ (156) structure. The A-LO and S-LO modes, in turn, lead to even lower symmetric structures with SG $Amm2$ (38) and $Cm$ (6), respectively. 

\begin{figure}[t]
    \includegraphics[width=\columnwidth]{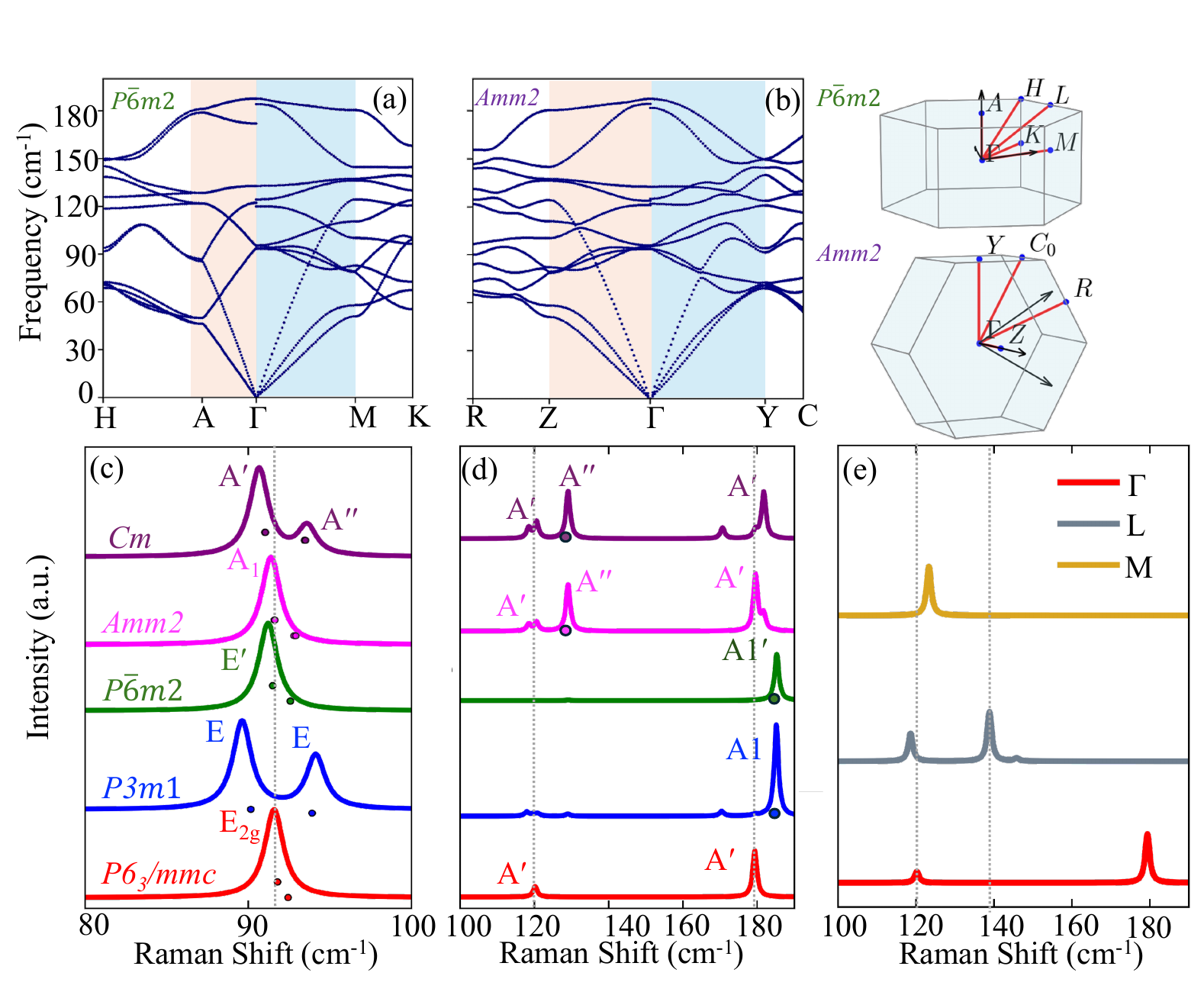}
    \caption{(a, b) Phonon band structures for two distinct TO ($P\bar{6}m2$) and LO ($Amm2$) distortion of MnTe. The atoms are displaced along the identified phonon modes to the tune of a mass-weighted mode amplitude of 0.1 ~\AA$\cdot$amu$^{1/2}$. Their corresponding BZs are shown on the right panel. (c, d) LR-corrected Raman spectra of pristine MnTe with all the plausible distorted analogues for two separate frequency ranges 80 - 100 cm$^{-1}$, and 100 - 190 cm$^{-1}$, respectively. The Raman peak positions without LR interaction are represented through colored dots in the spectra. (e) The off-center phonon contribution to the Raman spectra emerging out of the M and L centered phonons and are compared 
    to that of the zone centered phonons.}
    \label{raman}
\end{figure}

The LR corrected phonon band structures for two representative LO and TO distortions, $P\bar{6}m2$ and $Amm2$, are shown in Fig. \ref{raman}(a, b), remaining two are shown in Fig. S3(a,b). The $P\bar{6}m2$ distortion produces only minor changes to the phonon bands of the pristine MnTe. On the other hand, when compared to the dispersion along the transverse $\Gamma$-A and longitudinal $\Gamma$-M for the pristine MnTe (see Fig. \ref{SR+LR}), the $Amm2$ distorted structure with a different BZ (Fig. \ref{raman}(b)) exhibits contrasting phonon dispersion along the transverse $\Gamma$-Z and longitudinal $\Gamma$-Y directions (see the highlight in Fig. \ref{raman}(a)). As observed from Fig. \ref{raman}(b), $Amm2$ type distortion reduces the directional disparity of the modes around zone-center. The band structures corresponding to $P3m1$ and $Cm$ distorted structure are nearly identical to the case of $P\bar6m2$ and $Amm2$, respectively and hence are not shown.

To comprehend the overall changes in the vibrational landscape of MnTe, we calculated the first-order Raman spectra of pristine and distorted MnTe by taking into account the NAC. The detailed methodology for calculating Raman intensities and constructing Raman spectral envelope from LR corrected phonon frequencies and eigenvectors is provided in SM-V. The zone-center Raman spectra are presented in Fig. \ref{raman}(c) and (d) for two separate frequency ranges of 80 - 100 cm$^{-1}$ and 100 - 190 cm$^{-1}$, respectively. The positions of the peaks without LR interaction are also indicated by colored dots. We emphasize that, several experimental studies\cite{azam2026origin, bhardwaj2026nature175180cm1raman} have reported four distinct Raman features below 200 cm$^{-1}$, centered around 90 cm$^{-1}$, 122 cm$^{-1}$, 140 cm$^{-1}$, and 175 cm$^{-1}$. Although theoretical calculations ignoring the LR dipole-dipole interactions have correctly reproduced the 90 cm$^{-1}$ peak\cite{wu2026domains, uykur2026revisiting}, they have failed to do so for the remaining three. 
%Therefore, the microscopic origin of the other three peaks has remained far from resolved. 
Various mechanisms, including distortion-mediated and plasmon-related effects, have been proposed to account for the higher frequency Raman features in MnTe (see the first section). Raman spectra for a MnTe$_2$, appearing as an impurity phase in MnTe is also calculated, see SM-VI. In this study, by incorporating the LR dipole–dipole interaction and by resolving the associated splitting in previously degenerate LO and TO modes, we reproduce all four experimentally observed Raman peaks under appropriate structural and vibrational conditions.

The prominent Raman peak near 90 cm$^{-1}$ in Fig. \ref{raman}(c) corresponds to the robust E$_{2g}$ type vibration of pristine MnTe. This mode remains the most intense Raman peak irrespective of the structure being pristine or distorted. However, the distortions $P\bar{6}m2$ and $Cm$ cause a small split of this mode by $\sim$4 cm$^{-1}$, while preserving its characteristic vibrational pattern. Ukyur \textit{et al.} in their experimental studies have reported the presence of close-spaced double peaks corresponding to this E$_{2g}$ mode\cite{uykur2026revisiting}.

As observed in Fig. \ref{raman}(d), for pristine MnTe, the peak near 120 cm$^{-1}$ originates from the LO branch along $\Gamma$-L with A$^{\prime}$ symmetry and is revealed by the LR correction (see inset of Fig. \ref{SR+LR}). Without the correction, this mode had the A$_{2u}$ symmetry and is Raman inactive. 
%The LR interaction shifts this branch (as discussed earlier through Fig. 3) and makes this otherwise hidden mode Raman active. 
The Raman response also depends strongly on the distortion, such as, the TO-type distortions ($P\bar{6}m2$, $P3m1$) suppress the 120 cm$^{-1}$ feature, whereas the LO-type distortions ($Amm2$, $Cm$) intensify it. LO type distortions also give rise to another peak at 129 cm$^{-1}$ which was also silent (E$_{1u}$) in pristine MnTe.

The LR dipole-dipole interactions introduce a new Raman peak in pristine MnTe that appears at 179 cm$^{-1}$ (see Fig. \ref{raman}(d)). We traced its origin to the splitting of the S-LO phonon along the $\Gamma$-L direction (see Fig. \ref{SR+LR}) and we attribute it as one of the reasons for experimentally observing a peak around 175 cm$^{-1}$. The same feature is retained in the both LO-distorted structures, but disappears for the TO-type distortions. Instead, the TO-distorted structures develop another A$_1$-type Raman mode near 185 cm$^{-1}$ both with and without LR correction. This is consistent with the discussions made by Wu \textit{et al.}, where the authors have pinned down the highest TO mode to be Raman active under this specific (S-TO distortion) as the symmetry of this mode changes from B$_{2u}$ to A$_1^{\prime}$\cite{wu2026domains}. Furthermore, S-LO distortion corresponding to $Cm$ symmetry shows a Raman active mode at 170 cm$^{-1}$ (see the upper panel Fig. 4(d)). %\textcolor{red}{(How are we defining S-LO/TO and A-LO/TO modes?)}

Next, we discuss the experimental Raman peak $\sim$ 140 cm$^{-1}$. Previous reports have attributed it to the existence of zone off-center Raman modes \cite{wu2026domains, uykur2026revisiting}. From our calculated Raman spectra, we indeed found a Raman peak at 140 cm$^{-1}$ (see Fig. \ref{raman}(e)) and they originate from zone boundary L-phonons. Since this peak does not arise from zone-center phonons, but instead comes to existence from off-center contributions at L and M, the experimental detection of this particular peak may indicate the activation of phonons through higher-order Raman processes\cite{padmanabhan2022interlayer, guntherodt1979zone}.
In addition, we see two new peaks around 120 cm$^{-1}$ arising from L and M phonons. Therefore, 120 cm$^{-1}$ peak is collectively formed by zone-center and zone-boundary phonons. 
%The origin of the experimentally observed feature near \(140~\mathrm{cm}^{-1}\) is different.

In summary and conclusion, we provide the missing theoretical explanation of the experimentally observed landscape of the Raman response of MnTe. We demonstrate that the LR dipole-dipole interaction is the key to accurately calculate the phonon band structure of this altermagnetic insulator and it is due to the fact that MnTe possess large directional-dependent Born effective charges. The LR corrections to the phonon band structure introduce additional modes due to splitting, and subsequently some of the silent modes give rise to Raman Active modes. Our study reveals Raman active modes around 120 cm$^{-1}$, 140 cm$^{-1}$, and 179 cm$^{-1}$ which are in addition to the existing 90 cm$^{-1}$ peak, which is reproduced even without LR correction. Further analysis shows that the 140 cm$^{-1}$ peak originates from the zone-boundary phonons, while the 120 cm$^{-1}$ peak has contribution from both zone-center and zone boundary phonons. The highly debated Raman peak around 175 cm$^{-1}$ seems to have many origins, including LR effects in pristine MnTe, and LO distortions. The Robust Raman peak around 175 cm$^{-1}$ due to the presence of an impurity MnTe$_2$ phase can be considered as an experimental anomaly. %We note that, MnTe$_2$ phases well-established $T_g$ and $A_g$ Raman peaks around 175 cm$^{-1}$\cite{palasyuk2026paramagnetic}.} 
\section{Acknowledgement}
SJ thanks IIT Madras for providing fellowship, HPCE, and Kamadhenu cluster for the required computational support.  
%\putbib[ref]
%\end{bibunit}
\clearpage
%\begin{bibunit}[apsrev4-2]

% ==================================
% SUPPLEMENTARY MATERIAL
% ==================================

\clearpage

\setcounter{section}{0}
\setcounter{subsection}{0}

\renewcommand{\thesection}{S\arabic{section}}
\renewcommand{\thesubsection}{S\arabic{section}.\arabic{subsection}}

\setcounter{equation}{0}
\renewcommand{\theequation}{S\arabic{equation}}

\setcounter{figure}{0}
\renewcommand{\thefigure}{S\arabic{figure}}

\setcounter{table}{0}
\renewcommand{\thetable}{S\arabic{table}}

\renewcommand{\theHsection}{supp.\arabic{section}}
\renewcommand{\theHequation}{supp.\arabic{equation}}
\renewcommand{\theHfigure}{supp.\arabic{figure}}
\renewcommand{\theHtable}{supp.\arabic{table}}
\section*{Supplementary Material}
\section{I: Supercell convergence test}
\begin{figure}[bhtp]
    \centering
    \includegraphics[width=\columnwidth]{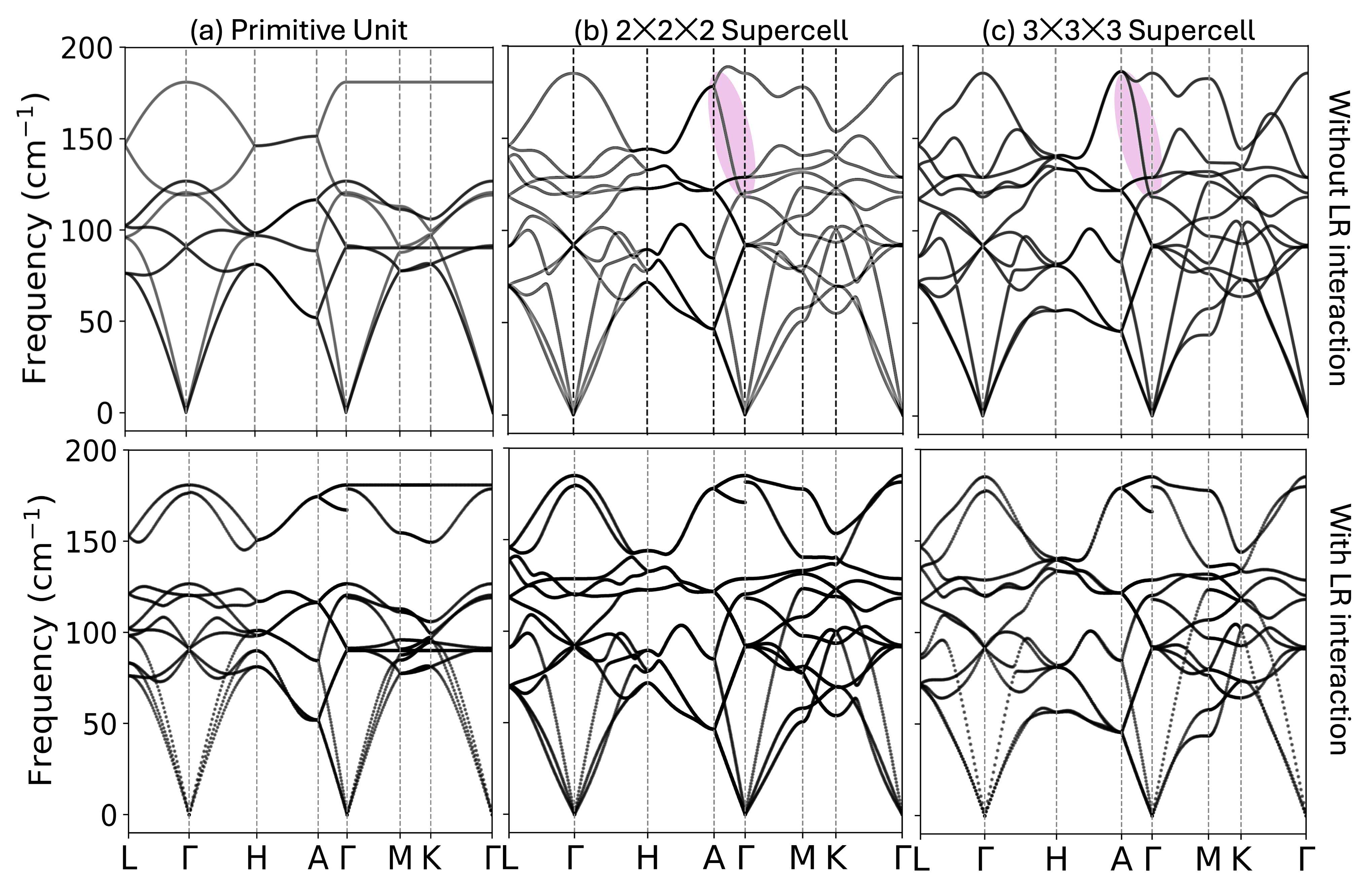}
    \caption{Phonon band structures of MnTe with varying supercell size. The upper and lower panels show the band structures with and without LR-corrections respectively. The long-range dipole-dipole contribution to the dynamical matrix, resulted in improved convergence with a 2$\times$2$\times$2 supercell, which is the minimum supercell size to retain all the operational symmetries for the space group.}
    \label{super_conv}
\end{figure}

\section{II: Non analytical correction in long wavelength limit}

The Born effective charge tensor ($Z^*$) quantifies the change in macroscopic polarization induced by an infinitesimal atomic displacement and is defined as
$Z_{\kappa,\alpha\beta}^{*}
= \frac{\Omega}{e}
\left.
\frac{\partial P_{\alpha}}
{\partial u_{\kappa\beta}}
\right|_{\mathbf{E}=0}$,
where $\Omega$ is the unit-cell volume, $e$ is the elementary charge, and $P_{\alpha}$ denotes the $\alpha$ component of the macroscopic polarization.
Here, $u_{\kappa\beta}$ represents the displacement of atom $\kappa$ along the direction $\beta$, while $\alpha$ specifies the polarization
direction ($\alpha, \beta$ = $x$, $y$, or $z$) (in zero external field). Thus, $Z^*$ characterizes the coupling between ionic displacements and electronic response as macroscopic polarization. 

For a phonon characterized by the wave vector $\mathbf{q}$, a nonzero projection of the Born effective charge, $\mathbf{q}\cdot\mathbf{Z}_{n}^{*}\neq 0$, gives rise to a long-range electric field associated with lattice vibration. In the long-wavelength limit ($\mathbf{q}\rightarrow 0$), this field introduces a nonanalytic, direction-dependent dipole-dipole contribution to the dynamical matrix, which is screened by the high frequency dielectric tensor ($\boldsymbol{\epsilon}_{\infty}$) of the material. This non-analytical correction (NAC) to the dynamical matrix, therefore, can be written as\cite{togo2022mode, gonze1997dynamicalsi,coiana2024breakdown}
\begin{equation}
D^{\mathrm{NA}}_{\kappa\alpha,\kappa'\beta}(\mathbf{q}\rightarrow0)
=
\frac{4\pi e^2}{\Omega}
\frac{
\left[\mathbf{q}\cdot Z^{*}_{\kappa}\right]_{\alpha}
\left[\mathbf{q}\cdot Z^{*}_{\kappa'}\right]_{\beta}
}{
\sqrt{M_{\kappa}M_{\kappa'}}
\left(\mathbf{q}\cdot
\boldsymbol{\epsilon}_{\infty}\cdot\mathbf{q}\right)
},
\end{equation}
where $M_{\kappa}$ and $M_{\kappa'}$ are the atomic masses corresponding to the dynamic charges, $Z^{*}_{\kappa}$ and $Z^{*}_{\kappa'}$. The total dynamical matrix around the point $\Gamma$ was, therefore, obtained by
\begin{equation}
D(\mathbf{q}) = D^{\mathrm{A}}(\mathbf{q}) + D^{\mathrm{NAC}}(\mathbf{q}\rightarrow0),
\end{equation}
where $D^{\mathrm{A}}$ represents the analytic part obtained from the short-range interatomic force constants.

We have obtained the $Z^{*}$ and ${\epsilon}_{\infty}$ tensors from first-principles calculations for the $\alpha$-MnTe system. Incorporating these quantities into $D(\bold{q})$ accounts for the direction-dependent modifications to the phonon dispersion spectra, as shown in Fig. 2(a). The optical phonon branches near the $\Gamma$ point experiences significant changes, including possible LO-LO and TO-TO splitting in MnTe. The magnitude of this splitting depends on the ratio of $Z^{*}$ and ${\epsilon}_{\infty}$, and the propagation direction, $\bold{q}$. %Away from the $\Gamma$ point, the NAC contribution diminishes, and the corrected and uncorrected phonon dispersion spectra are therefore expected to approach one another.\\
Hence, we calculate the mode effective charge by projecting $Z^{*}$ onto the corresponding phonon mode, $e_{\kappa,m}$. For this particular vibrational eigenvector, represented by the unit vector $\mathbf{q}_m$, the effective charge of the mode is expressed as\cite{molina2011phonons, srinivasan2025understanding},
\begin{equation}
Z_{m,\kappa}(\mathbf{q}) =
\sum_{\alpha,\beta}
q_m Z_{\kappa,\alpha\beta}^{*}
\label{eq:mode_effective_charge}
\end{equation}
A nonzero value of \(Z_m(\mathbf{q})\) for each sublattice vibration indicates that the corresponding phonon mode can create long-range order in the system, eventually creating macroscopic polarization.

As shown in Table I, $Z^{*}$ of Mn and Te exhibit nearly equal magnitudes but opposite signs, with the in-plane components varying from approximately 3.528 to 2.706, and the out-of-plane components from 2.735 to 2.571 with the increase in $U$. ${\epsilon}_{\infty}$, is also anisotropic across the considered $U$ range, with in-plane components varying between 29.52 - 10.07, and the out-of-plane component between 28.92 - 10.32.

\begin{table*}[t]
\centering
\footnotesize
%\resizebox{\columnwidth}{!}{%
\begin{tabular}{c|c}
\hline
Atom & Born effective charge $Z^*$ \\
\hline
Mn &
$\displaystyle
\begin{pmatrix}
3.528 \ge Z^* \ge 2.706 & 0.00 & 0.00 \\
0.00 & 3.528 \ge Z^* \ge 2.706 & 0.00 \\
0.00 & 0.00 & 2.735 \ge Z^* \ge 2.571
\end{pmatrix}$
\\[12pt]
Te &
$\displaystyle
\begin{pmatrix}
-3.528 \le Z^* \le -2.7057 & 0.00 & 0.00 \\
0.00 & -3.528 \le Z^* \le -2.7057 & 0.00 \\
0.00 & 0.00 & -2.735 \le Z^* \le -2.571
\end{pmatrix}$
\\
\hline
\multicolumn{2}{c}{} \\[-8pt]
\multicolumn{2}{c}{
$\displaystyle
\epsilon_\infty =
\begin{pmatrix}
29.52 \ge \epsilon \ge 10.07 & 0.00 & 0.00 \\
0.00 & 29.52 \ge \epsilon \ge 10.07 & 0.00 \\
0.00 & 0.00 & 28.92 \ge \epsilon \ge10.32
\end{pmatrix}$}
\\
\hline
\end{tabular}
%}
\caption{Range of Born effective charge, $Z^*$, and dielectric tensor, $\epsilon_\infty$ for U values from 0 to 7 eV.}
\label{tab-born}
\end{table*}

\begin{table*}[t]
\footnotesize
\setlength{\tabcolsep}{3.5pt}
\renewcommand{\arraystretch}{1.12}
%\resizebox{\columnwidth}{!}{%
\begin{tabular}{cc cc cc}
\toprule
\multicolumn{2}{c}{$\omega$ (cm$^{-1}$)} &
\multicolumn{2}{c}{$\mathbf{Z}_{\rm Mn}^{*}$} &
\multicolumn{2}{c}{$\mathbf{Z}_{\rm Te}^{*}$} \\
\cmidrule(lr){1-2}
\cmidrule(lr){3-4}
\cmidrule(lr){5-6}
Before LR & After LR &
Before LR & After LR &
Before LR & After LR \\
\midrule
\multirow{2}{*}{128.74}
& \shortstack{181.56\\ ($\Gamma$-M)}
& \multirow{2}{*}{\shortstack{$(3.27,0,0)$; \\ $(0,3.27,0)$}}
& $(-2.84,-1.64,0)$
& \multirow{2}{*}{\shortstack{$(2.14,0,0)$; \\ $(0,2.14,0)$}}
& $(-1.86,-1.07,0)$ \\
& \shortstack{128.74\\ ($\Gamma$-M)}
& & $(-1.64,2.84,0)$
& & $(-1.07,1.86,0)$ \\
\midrule
\multirow{2}{*}{118.13}
& \shortstack{170.42\\ ($\Gamma$-A)}
& \multirow{2}{*}{$(0,0,3.16)$}
& $(0,0,3.17)$
& \multirow{2}{*}{$(0,0,2.08)$}
& $(0,0,2.08)$ \\
& \shortstack{118.13\\ ($\Gamma$-M)}
& & $(0,0,3.16)$
& & $(0,0,2.08)$ \\
\midrule
\multirow{2}{*}{128.74}
& \shortstack{179.36\\ ($\Gamma$-L)}
& \multirow{2}{*}{\shortstack{$(3.27,0,0)$; \\ $(0,3.27,0)$}}
& $(-2.59,-1.49,1.29)$
& \multirow{2}{*}{\shortstack{$(2.14,0,0)$; \\ $(0,2.14,0)$}}
& $(-1.70,-0.98,0.85)$ \\
& \shortstack{128.74\\ ($\Gamma$-L)}
& & $(-1.64,2.84,0)$
& & $(-1.07,1.86,0)$ \\
\midrule
118.13
& \shortstack{120.22\\ ($\Gamma$-L)}
& $(0,0,3.16)$
& $(1.16,0.67,2.89)$
& $(0,0,2.08)$
& $(0.76,0.44,1.90)$ \\
\bottomrule
\end{tabular}%
%}
\caption{Phonon frequencies and atom resolved mode effective
charges ($Z_{m,\kappa}(\mathbf{q})$) before and after the inclusion of LR interaction.}
\label{tab:LR}
\end{table*}
In MnTe, the direction dependent lifting of optical phonon degeneracy near the $\Gamma$ point can be understood through the calculation of mode effective charges, $Z_m(q)$, summarized in Table \ref{tab:LR}. Before including LR interactions, the optical modes at 128.74 cm$^{-1}$ are doubly degenerate, with linear polarizations arising from the relative displacement of the Mn and Te sublattices. The effective charges in the corresponding mode are $Z_{\mathrm{Mn}}=(3.27,0,0)$ and $Z_{\mathrm{Te}}=(2.14,0,0)$. Their parallel orientation indicates that the cumulative contributions of Mn and Te reinforce the induced macroscopic polarization despite their opposite $Z^{*}$ (see Table \ref{tab-born}). Upon inclusion of the LR interaction, the split observed for a degenerate mode $m$, can be quantitatively estimated as,
\begin{equation}
\Delta\omega_m^2 =
\frac{e^2}{\epsilon_0\Omega}
\frac{
|\mathbf{q}\cdot Z_m|^2
}{
\mathbf{q}\cdot\epsilon_\infty\cdot\mathbf{q}
}
\label{LO_TO_splitting}
\end{equation}
$\mathcal{Z}_m$ is the mass weighted $Z_{m,\kappa}(\mathbf{q})$.
Using the atomic masses of Mn (54.938 amu) and Te (127.60 amu), the mass weighted mode effective charge for propagation along $q=(1,0,0)$ is,
$Z_{xx}$ =$\frac{3.27}{\sqrt{54.938}} +
\frac{2.14}{\sqrt{127.60}}$$\approx$ 0.631.
For the primitive cell of MnTe, $\Omega=103.9603$ \AA $^3$ and $\epsilon_{\infty,xx}=11.5224$(see Table \ref{tab-born}), the squared frequency difference ($\tilde{\nu}$) becomes
\begin{equation}
\Delta\tilde{\nu}^2 = 41078.17|Z_{xx}|^2.
\end{equation}
Substituting the calculated $Z_{xx}$ gives $\tilde{\nu}_{new}\approx181.4$ cm$^{-1}$ from the initial value of 128.74 cm$^{-1}$, corresponding to an LO-LO split of approximately 52.7 cm$^{-1}$. This agrees well with the LR corrected phonon spectra, where the mode shifts from 128.74 to 181.56 cm$^{-1}$ (see Table \ref{tab:LR}).

Similarly, along $\Gamma$-L ($\mathbf{q}=(1,0,1)$), the mode shift occurs at 179.36 from 128.74 cm$^{-1}$. The corresponding $Z_{m,\kappa}(\mathbf{q})$ also change orientation and become plane-polarized as a result of this correction, indicating a redistribution of the eigenvector character (amplitude $|Z_{m,\kappa}(\mathbf{q})|$ remains the same). Similar kind of behaviour is also observed along the in-plane $\Gamma$-K direction ($\mathbf{q}=(1,1,0)$) as well. Thus, the LR interaction modifies both the phonon frequencies and the polarization character of the optical modes in MnTe, depending on the propagation direction. Furthermore, along the $\Gamma$-A direction, the optical mode at 118.13 cm$^{-1}$ shifts to 170.42 cm$^{-1}$ upon inclusion of the LR interaction. The atom resolved mode effective charges remain unchanged for the out-of-plane direction.
\section{III: Energy landscape with and without LR interaction}
\begin{figure}[bhtp]
    \centering
    \includegraphics[width=\columnwidth]{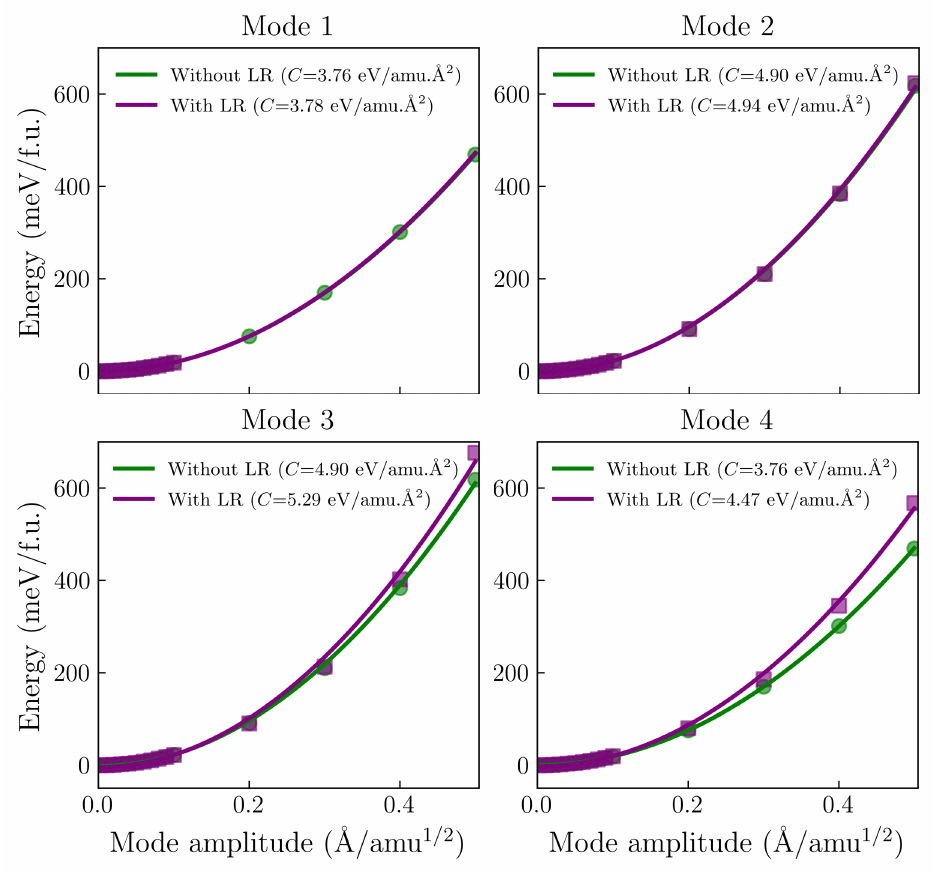}
    \caption{Relative energy change as a function of phonon mode amplitude before and after the inclusion of LR interactions for the four shifted modes labeled 1-4 [see also Fig. 2(a)]. The energy profiles are fitted using a quadratic function to extract the corresponding harmonic force constants ($C$). Notably, modes 3 and 4, associated with phonon propagation along the $\Gamma$–L direction, exhibit a significant increase in $C$ upon inclusion of the LR correction.}
    \label{fc_fit}
\end{figure}

For a phonon mode $m$, characterized by the eigenvector $e_{\kappa\alpha}^{m}$, the atomic displacement is expressed as,
\begin{equation}
\Delta R_{\kappa\alpha}^{m}
= u \frac{e_{\kappa\alpha}^{m}}{\sqrt{M_\kappa}},
\end{equation}
where $u$ is the mass-weighted mode amplitude and
$M_\kappa$ is the mass of atom $\kappa$ (phase factor, $\phi$ = 0 for zone-center vibrations). The potential energy along this vibrational mode
can be expanded as,
\begin{equation}
U_m(u) = U_0
+ \left(\frac{dU_m}{du}\right)_0 u
+ \frac{1}{2}
\left(\frac{d^2U_m}{du^2}\right)_0 u^2
+ \mathcal{O}(u^3).
\end{equation}
At equilibrium, the linear term vanishes. Within the harmonic approximation,
\begin{equation}
\Delta U_m(u)
= U_m(u)-U_0
= \frac{1}{2}C_m u^2,
\qquad
C_m = \left(\frac{d^2U_m}{du^2}\right)_0.
\end{equation}
Following this formalism, we employed a mode dependent frozen-phonon approach, in which atomic configurations are systematically distorted along the corresponding phonon eigenvectors. The distorted atomic positions are then obtained as,
\begin{equation}
R_{\kappa\alpha}^{m}(u) = R_{\kappa\alpha}^{0} + \Delta R_{\kappa\alpha}^{m}(u),
\label{eq:distorted_positions}
\end{equation}
where $R_{\kappa\alpha}^{0}$ denotes the equilibrium atomic position. For each selected mode, a series of distorted structures is generated by varying $u$ while keeping the eigenvector fixed. The corresponding total energies are evaluated without further atomic relaxation, allowing the potential energy surface to be mapped along the chosen vibrational coordinate. We quantify the effect of LR interactions on the vibrational potential of MnTe and examined the relative potential energy (U) as a function of newly introduced phonon mode amplitude. Within the harmonic approximation, the energy retains its parabolic dependence on the mode amplitude, expressed as $\Delta U=\frac{1}{2}Cu^2$. We focus on four modes (1-4) highlighted in Fig. 2 of the main manuscript, which exhibit distinct responses to the LR correction. Owing to the symmetry of the harmonic potential, only positive mode amplitudes are shown. As illustrated in Fig. 2, modes 1 and 2, associated with vibrations along the $\Gamma$-A and $\Gamma$-M directions, respectively, exhibit negligible changes in their energy profiles and force constants upon inclusion of the LR interaction. In contrast, mode 3, propagating along $\Gamma$-L, shows a pronounced increase in the curvature of the potential energy, indicating enhanced vibrational stiffness. Mode 4 exhibits the largest increase in the fitted force constant, from 3.76 to 4.47 eV/amu.\AA$^{2}$. These results helped us quantifying the overall effects of LR correction on the vibrational restoring forces in a mode and direction dependent manner, leading to selective phonon hardening in MnTe around the zone-center.
\section{IV. Group-theoretical Analysis of Phonon Modes}
The original group of MnTe points is $D_{6h}$ that contains 24 symmetry operations. At $\Gamma$, all 24 symmetry operations are respected, but away from the the center, certain symmetry operations are violated. The reduction of these symmetry operations gives rise to several subgroups. After the inclusion of LR, there are four new direction dependent phonon modes around the center whose irreducible representations depend on the path symmetry along that direction, as seen in Fig. \ref{SR+LR}. As a result, along $\Gamma$-A path the $C_{6v}$, $\Gamma$-M path $C_{2v}$, and $\Gamma$-L $C_s$ point group symmetry is maintained.
These groups contain 12, 4, and 2 symmetry operations, respectively. %, and form the subgroup chain, $D_{6h} \supset C_{6v} \supset C_{2v} \supset C_s$.
\vspace{0.3cm}
\begin{table}[htbp]
\centering
\footnotesize
\caption{Character table of the $C_{6v}$ point group.}
\begin{tabular}{c c c c c c c l}
\hline\hline
$C_{6v}$ & $E$ & $2C_6$ & $2C_3$ &
$C_2$ & $3\sigma_v$ & $3\sigma_d$ &
Basis functions \\
\hline
$A_1$ & 1 &  1 &  1 &  1 &  1 &  1 &
$z;\ x^2+y^2,z^2$ \\
$A_2$ & 1 &  1 &  1 &  1 & -1 & -1 &
$R_z$ \\
$B_1$ & 1 & -1 &  1 & -1 &  1 & -1 &
-- \\
$B_2$ & 1 & -1 &  1 & -1 & -1 &  1 &
-- \\
$E_1$ & 2 &  1 & -1 & -2 &  0 &  0 &
$(x,y);\ (xz,yz)$ \\
$E_2$ & 2 & -1 & -1 &  2 &  0 &  0 &
$(x^2-y^2,xy)$ \\
\hline\hline
\end{tabular}
\label{tab:C6v}
\end{table}

The eigenvector for Mode 1 at 170.42 cm$^{-1}$ along $\Gamma$-A direction signifies an out of plane optical vibration of two atomic sublattices in opposite directions and is of A$_1$ character with Raman tensor, R(A$_1$) as, 
\begin{equation}
R(A_1)=
\begin{pmatrix}
a&0&0\\
0&a&0\\
0&0&b
\end{pmatrix},
\end{equation}
\vspace{0.3cm}
\begin{table}[htbp]
\centering
\footnotesize
\caption{Character table of the $C_{2v}$ point group.}
\begin{tabular}{c c c c c l}
\hline\hline
$C_{2v}$ & $E$ & $C_2(z)$ &
$\sigma_{xz}$ & $\sigma_{yz}$ & Basis functions \\
\hline
$A_1$ & 1 &  1 &  1 &  1 &
$z;\ x^2,y^2,z^2$ \\
$A_2$ & 1 &  1 & -1 & -1 &
$xy;\ R_z$ \\
$B_1$ & 1 & -1 &  1 & -1 &
$x;\ xz$ \\
$B_2$ & 1 & -1 & -1 &  1 &
$y;\ yz$ \\
\hline\hline
\end{tabular}
\label{tab:C2v}
\end{table}

Along $\Gamma$-M, Mode 2 at 181.56 cm$^{-1}$ behaves as a LO mode where atomic displacements are strictly aligned at approximately 30$^\circ$ to the $x$ axis, with Mn and Te moving oppositely in-plane. This mode transforms according to the totally symmetric $A_1$ irreducible representation of the $C_{2v}$ group.
with R(A$_1$) as,
\begin{equation}
R(A_1)=
\begin{pmatrix}
a&0&0\\
0&b&0\\
0&0&c
\end{pmatrix}
\end{equation}
\vspace{0.3cm}
\begin{table}[htbp]
\centering
\caption{Character table of the $C_s$ point group.}
\begin{tabular}{c c c l}
\hline\hline
$C_s$ & $E$ & $\sigma_{xz}$ & Basis functions \\
\hline
$A'$  & 1 &  1 & $x,z;\ x^2,y^2,z^2,xz$ \\
$A''$ & 1 & -1 & $y;\ xy,yz$ \\
\hline\hline
\end{tabular}
\label{tab:Cs}
\end{table}

In both Modes 3 and 4, at 179.36 cm$^{-1}$ and 120.22 cm$^{-1}$, respectively, along $\Gamma$–L, the Mn and Te sublattices exhibit oppositely directed optical vibrations with coupled in-plane and out-of-plane components. For both modes, the in-plane displacement vectors are oriented at 30$^\circ$ to the $x$-axis. These modes transform according to the irreducible representation $A^{\prime}$ of $C_s$ with the corresponding Raman tensor, R(A$^{\prime}$) as,
\begin{equation}
R(A')=
\begin{pmatrix}
a&0&d\\
0&b&0\\
d&0&c
\end{pmatrix}
\end{equation}

All four modes lie close to 175 cm$^{-1}$ and have only diagonal Raman tensor components, making them observable in parallel polarization but absent in cross-polarization. This is consistent with the nature of the experimentally detected reports mentioned earlier. 
\section{V: Distortion Design in \protect\NoCaseChange{MnTe}}
\begin{figure}[bhtp]
    \centering
\includegraphics[width=\columnwidth]{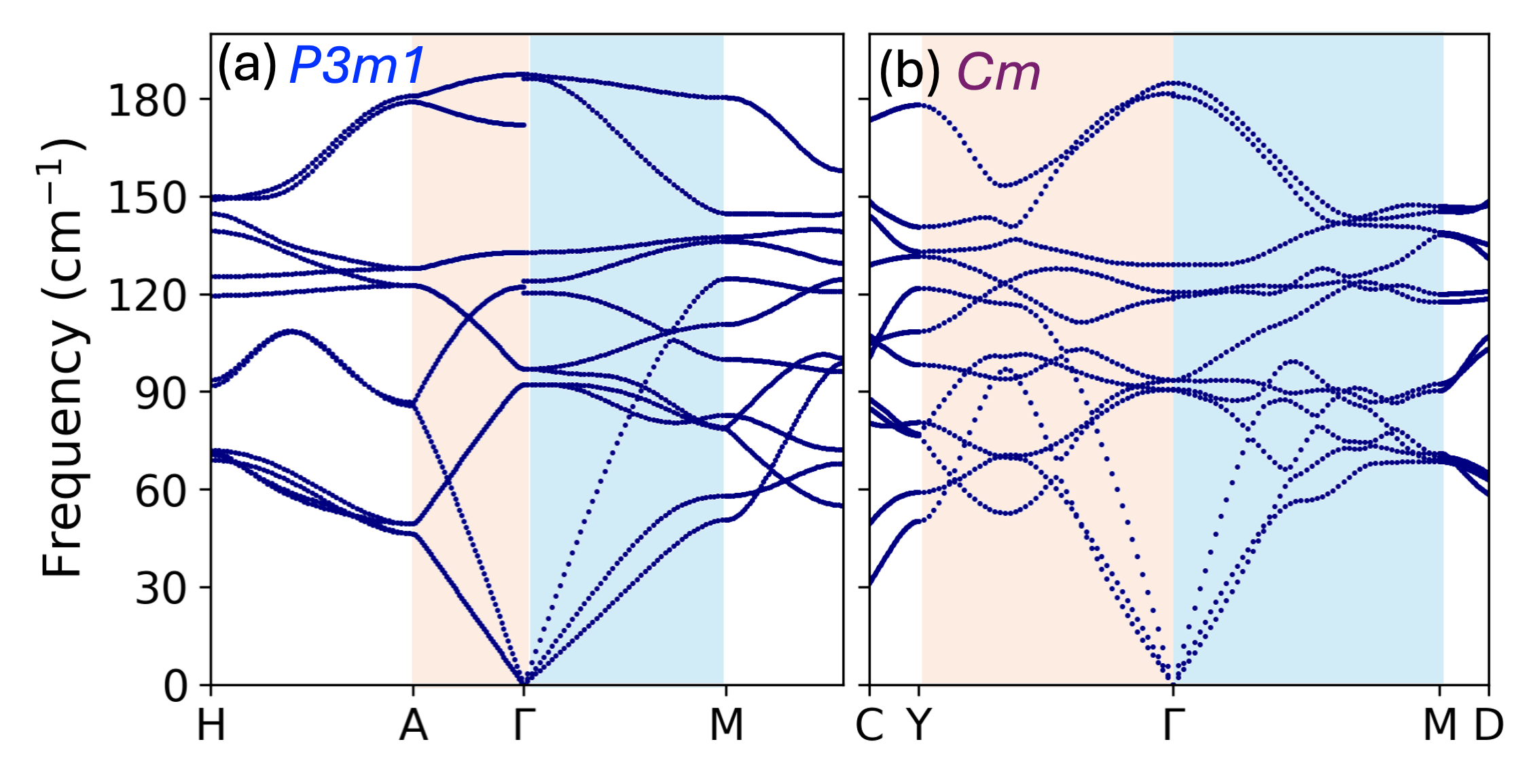}
    \caption{(a, b) Phonon band structures for the TO ($P3m1$) and LO ($Cm$) distortion of MnTe. The atoms are displaced along the identified modes to the tune of a mass weighted mode amplitude of 0.1 $\mathrm{\AA},\mathrm{amu}^{1/2}$.}  
%\end{subfigure}\begin{subfigure}\includegraphics[width=\columnwidth]{SDD.png}}\end{subfigure}
    \label{}
\end{figure}
To investigate the previously proposed mode dependent lattice distortions, we first evaluated the phonon group velocities, $v_{g,m}(q)=\nabla_q\omega_m(q)$, for the high-frequency optical branches along selected high-symmetry directions. Although the phonon group velocity does not directly represent the evolution of a structural distortion, it provides useful information about the direction of vibrational energy propagation. Based on the calculated $v_{g,m}(q)$ we identified four representative directions that exhibit appreciable group velocities, representing two predominantly transverse optical (TO) and two  longitudinal optical (LO) modes. Subsequently, mode-dependent frozen-phonon distortions were constructed using the corresponding phonon eigenvectors, with a fixed displacement amplitude of 0.1~\AA, to investigate the associated changes in phonon spectra. Two such distortion induced phonon band structures are given in Fig. 4(a) of the main manuscript and the other two are shown here in Fig. 3(a,b).
\section{VI: Details of Raman Spectra Calculation}
For subsequent estimation of Raman intensity, $\Gamma$-center phonon frequencies $\omega_j$ and eigenvectors $\mathbf{e}_j$ (mode, $j$ = 1, 12 for MnTe) are taken from the dynamical matrix included in LR interaction computed with Phonopy.
For each mode, two structures are generated by displacing every atom along $\pm\ 0.01$~\AA), as explained in SM-III.
The macroscopic static dielectric tensor ${\varepsilon}^{\pm}$ of each displaced structure is then computed using VASP. The Raman tensor of mode $j$ then calculated from the finite difference of $\varepsilon^+$, $\varepsilon^-$ with respect to the normal coordinate $u_j$,
\begin{equation}
  R^{(j)}_{\alpha\beta}
  = \frac{V}{4\pi}\,\frac{\partial \varepsilon_{\alpha\beta}}{\partial u_j}
  %\approx \frac{V}{4\pi}\,\mathcal{N}_j\,\frac{\varepsilon^{+}_{\alpha\beta}-\varepsilon^{-}_{\alpha\beta}}{2\Delta},
  \label{eq:ramantensor}
\end{equation}
where $V$ is the primitive cell volume. The calculated $R_{\alpha\beta}$ for each mode follows the symmetry constraints imposed by group theory, as discussed in SM-IV.
For each vibrational mode, $R_{\alpha\beta}$ can be reduced to the two rotational invariants, the isotropic component $\alpha_j$ and the anisotropic component $\beta_j^{2}$, where
\begin{align}
  \alpha_j &= \tfrac{1}{3}\big(R^{(j)}_{xx}+R^{(j)}_{yy}+R^{(j)}_{zz}\big), \label{eq:alpha}\\
    \beta_j^{2} &= \tfrac{1}{2}\Big[\big(R_{xx}-R_{yy}\big)^{2}
    +\big(R_{xx}-R_{zz}\big)^{2}+\big(R_{yy}-R_{zz}\big)^{2}\Big] \nonumber\\
  &\quad + 3\big(R_{xy}^{2}+R_{xz}^{2}+R_{yz}^{2}\big). \label{eq:beta}
\end{align}
The Raman activity of mode $j$ is then calculated as\cite{fonarivasp_raman, porezag1996infrared},
\begin{equation}
  I_j = 45\,\alpha_j^{2} + 7\,\beta_j^{2}.
  \label{eq:activity}
\end{equation}
This corresponds to the sum of the parallel and perpendicular components, $I_j = I_{\parallel}+I_{\perp}$, with $I_{\parallel}=45\alpha_j^2+4\beta_j^2$ and $I_{\perp}=3\beta_j^2$\cite{long2002raman}. Therefore, the Raman intensity calculated this way does not depend on the selective polarization environment.

\section{VII: Phonon spectra of \protect\NoCaseChange{MnTe$_2$}}
\begin{figure}[bhtp]
    \centering
    \includegraphics[width=\columnwidth]{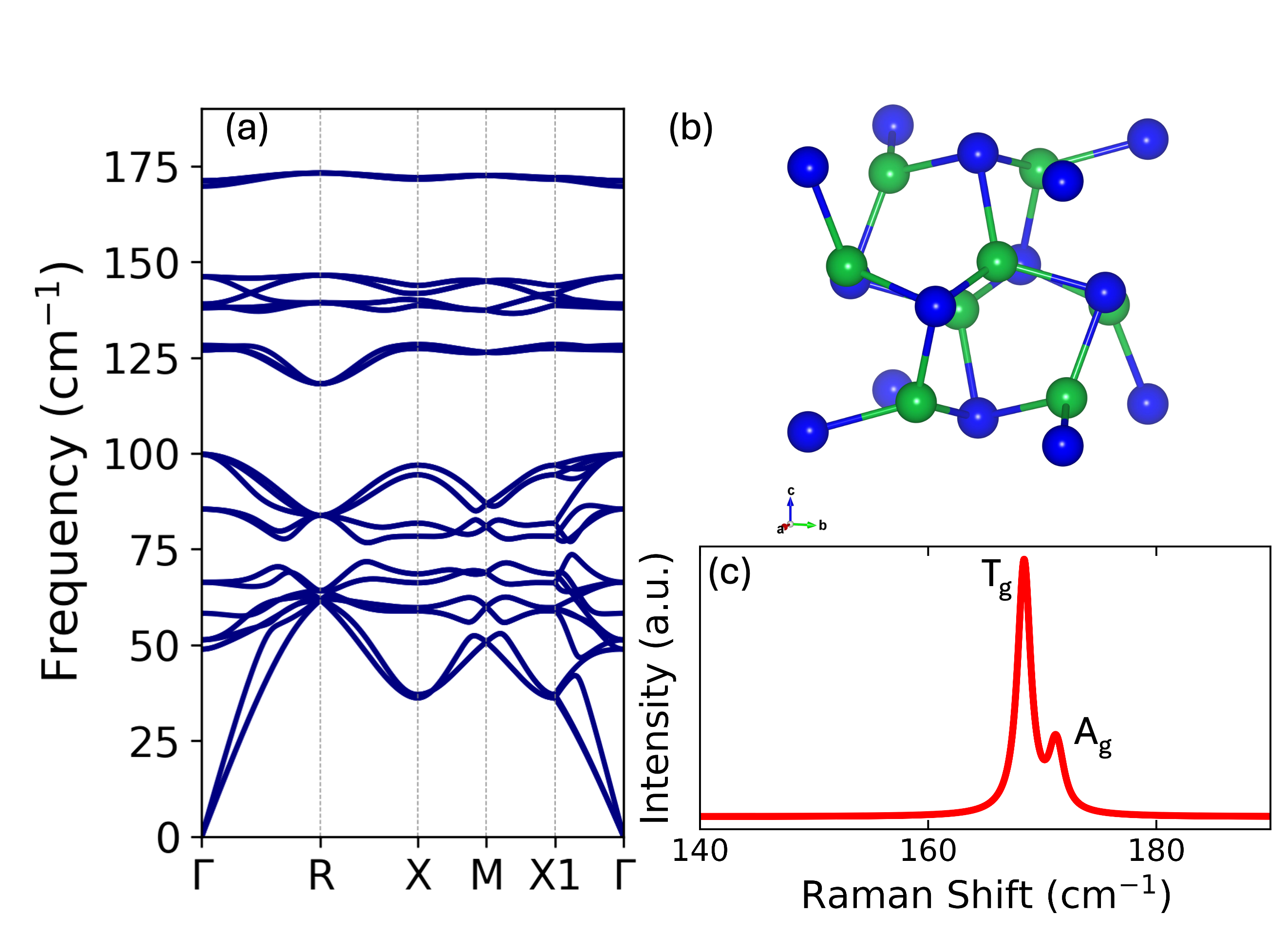}
    \caption{(a, b) Calculated phonon dispersion of $\alpha$-MnTe$_2$ and its corresponding bulk crystal structure, respectively. (c) Calculated Raman spectrum with vibrational mode assignments based on the $Pa\bar{3}$ (No. 205) space group symmetry. A prominent Raman peak is observed near 175 cm$^{-1}$, which may serve as a characteristic signature of $\alpha$-MnTe$_2$ in experimental MnTe samples containing Te-rich impurity phases.}
    \label{mnte2}
\end{figure}
In Fig. \ref{mnte2}, we have plotted the phonon band structure and the Raman response. The latter reproduces the well-established 175 cm$^{-1}$ peak. Some of the experimental studies on MnTe have reported this peak and they have attributed it to the impurity MnTe$_2$ phase.
\bibliography{ref}
%\putbib[ref]
%\end{bibunit}
\end{document}